\documentclass[epj]{svjour}
\usepackage{graphics}
\usepackage{graphicx}
\usepackage{dcolumn}
\usepackage{bm}
\usepackage{color}

\newcommand{\be}{\begin{equation}}
\newcommand{\ee}{\end{equation}}
\newcommand{\ba}{\begin{eqnarray}}
\newcommand{\ea}{\end{eqnarray}}
\newcommand{\bd}{\begin{displaymath}}
\newcommand{\ed}{\end{displaymath}}
\newcommand{\bea}{\begin{eqnarray}}
\newcommand{\eea}{\end{eqnarray}}
\newcommand{\di}{{\rm d}}
\renewcommand{\vec}[1]{\mbox{\boldmath$#1$}}

\begin{document}
\title{Fluid Dynamics Study of the $\Lambda$ 
Polarization for Au+Au Collisions at $\sqrt{s_{NN}}=200$ GeV}

\author{Yilong Xie\inst{1} \and Dujuan Wang\inst{2} 
\and Laszlo Pal Csernai\inst{3}}                     
%
%
\institute{School of Mathematics and Physics, China University of Geosciences(Wuhan), 
Lumo Road 388, 430074 Wuhan, China 
\and 
School of Science, Wuhan University of Technology, 
430070, Wuhan, China
\and
Institute of Physics and Technology, University of Bergen,
Allegaten 55, 5007 Bergen, Norway
}
\date{Received: date / Revised version: date}
%
\abstract{
With a Yang-Mills field, stratified shear flow initial state and a high resolution (3+1)D Particle-in-Cell Relativistic (PICR) hydrodynamic model, we calculate the $\Lambda$ polarization for peripheral Au+Au collisions at RHIC energy of $\sqrt{S_{NN}}=200$ GeV. The obtained longitudinal polarization in our model agrees with the experimental signature and the quadrupole structure on transverse momentum plane. It is found that the relativistic correction (2nd term), arising from expansion and from the time component of the thermal vorticity, plays a crucial role in our results. This term is changing the signature and exceeds the first term, arising from the classical vorticity. Finally, the global polarization in our model shows no significant dependence on rapidity, which agrees with the experimental data. It is also found that the second term flattens the sharp peak arising from the classical vorticity (1st term). 
} 
\maketitle
\section{Introduction}


In non-central relativistic heavy ion collisions, after penetrating each other, the Lorentz contracted nuclei will break down into quarks and gluons, forming the so called quark gluon plasma (QGP), which carries substantial amount of initial orbital angular momentum 
\cite{BPR2008,GCD2008}. 
The initial shear flow in the viscous QGP will lead to the rotation/vorticity and then the induced vorticity, via the spin-orbit interaction, will eventually give rise to the spin alignment of particles. The non-trivial local and global polarization, which is aligned with the initial angular momentum, was observed in many heavy ion collision experiments 
\cite{IA06,IS06,BIA07} 
and raising great interest among the physicists 
\cite{LW05a,HHW11,BGT07,BCDG13}.

The recent measurement of $\Lambda$ hyperon polarization in heavy ion collisions launched by the STAR collaboration at, RHIC has shown non-vanishing signals, even at 200 GeV
\cite{FirstSTAR,Nature,SecondSTAR}, 
and in low energy range, the signal could be as significant as 8\%. Presently, the experimental signals seem conform with the theoretical calculations and predictions, but still many puzzles remain in this field 
\cite{QMSTAR}.  

One of them is the sign problem: the longitudinal polarization on transverse momentum space shows flipped sign distribution with respect to (w.r.t) the sign distribution of vorticity induced by elliptic flow. Another problem is that the $\bar{\Lambda}$ polarization is significantly larger than $\Lambda$ polarization, e.g. by a factor of 4 at   $\sqrt{s_{NN}} = 7.7$ GeV, which heretofore has not been predicted and thereafter still has not been interpreted satisfactorily, by theory. There are proposals that the magnetic field induced by the spectators, plays a crucial role in splitting the $\bar{\Lambda}$ and $\Lambda$ polarizations, but the magnitude and the duration of the magnetic field emerging in early stage is still unclear
\cite{DH2012,MS2014,HX2018}. 
However, a recent work proposed that the magnetic field induced by the vortical baryonic QGP of participant system could last much longer time until freeze-out, although the magnitude is much smaller than that induced by charged spectators
\cite{GLW2019}.

More precise measurements of $\Lambda$ at the RHIC 200GeV Au+Au collisions reveal more disagreement between theory and experiment. E.g., the azimuthal distribution of longitudinal polarization shows flipped sign distribution compared to the hydro-simulated, z-directed polarization over transverse momentum space; the $y$-directed global polarization is larger in in-plane than in out-of-plane, which is opposite to the hydrodynamic simulations 
\cite{Xie2016,KB2017}, 
as well as transport model results
\cite{XLTW2018}; 
the global polarization shows no significant dependence on pseudo-rapidity and transverse momentum, while the simulations from a multiple phase transport (AMPT) model shows a normal distribution on pseudo-rapidity and decreasing with the transverse momentum.

In another words, the existing calculations and simulations with the approach developed by F. Becattini et al.
\cite{BCZG2013}, 
using the thermal vorticity at freeze-out to predict the final particle polarization, presently can  agree with experiments on the energy dependence behavior of the global polarization and the quadrupole structure of longitudinal polarization. When it comes to the differential measurements, there exists many discrepancies between them, as described above. This indicates some underlying misunderstandings among/inside the theory, simulation and experiments. E.g. the ref. 
\cite{ITS2019} 
proposed that the theoretically deduced polarization vector should be averaged over the momentum direction of emitted $\Lambda$ hyperons, to sample the experimental data. This results in a global polarization that is only proportional to the (spatial) thermal vorticity, differing from previous calculations where the temporal component of the relativistic vorticity was also included. As we can see later in this paper, this approach will have a global polarization that is larger in in-plane than out-of-plane, agreeing with the experimental data. Likewise, another ref. 
\cite{FKR2019}, 
also attempted to use the projected thermal vorticity as the source of spin polarization, and surprisingly they seem obtain a correct sign distribution of longitudinal polarization compared to experimental data.

However, up to now, there is still not a comprehensive calculation for the $\Lambda$ polarization at 200 GeV Au-Au collisions, which we could directly compare to experimental data.

Therefore, the main task of this paper is to simulate the 200 GeV Au+Au collisions, using a Yang-Mills flux-tube initial state and a high resolution (3+1)D Particle-in-Cell Relativistic (PICR) hydrodynamics model. We calculate comprehensively the local and global $\Lambda$ polarization (as function of different variables), that could be compared to the experimental data. By doing this, we expect to see directly how many discrepancies and agreements there exist among the theory, simulations and experimental results.

\section{Simulation and Polarization Vector}

The nucleus-nucleus impact in our initial state is divided into many slab-slab collisions, and Yang-Mills flux-tubes. These are assumed to form  streaks
\cite{M1,M2}. 
In this scenario, the initial state naturally generates longitudinal velocity shear flow, which when put into the subsequent high resolution (3+1)D Particle-in-Cell Relativistic (PICR) hydrodynamics model, will develop into substantial vorticity. Since our initial state+hydrodynamic model characterize the shear and vorticity in heavy ion collisions fairly well, its simulations to the $\Lambda$ polarization has achieved many successes.

As the first hydrodynamic model applied to the polarization study 
\cite{BCW2013}, 
this model was then widely used in our other previous works, e.g. refs. 
\cite{Xie2016,Xie2017}, 
and exhibited good descriptions to the vorticity and $\Lambda$ polarization. In the present work, we chose the modeling parameters as follows: the cell size  is $0.343^3$ fm$^3$, the time increment is 0.0423 fm/c, and the freeze-out (FZ) time is 4+4.91 fm/c (4 fm/c for the initial state's stopping time and 4.91 fm/c corresponds to the hydro-evolution time, which is similar to that in ref. 
\cite{BCW2013}).

\begin{figure}[ht] 
\begin{center}
      \includegraphics[width=8.5cm]{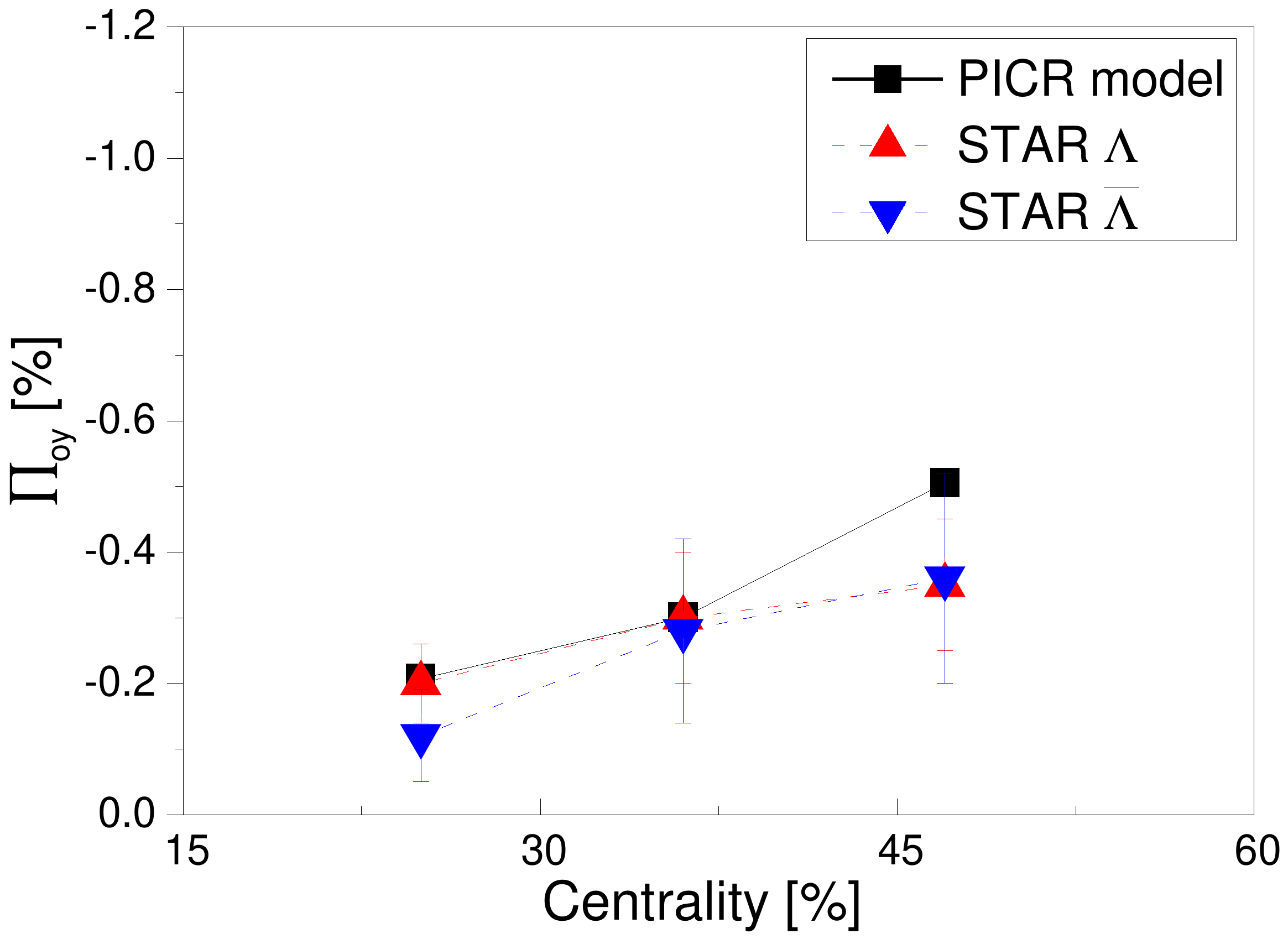}
\end{center}
\caption{
(Color online) The global polarization  for Au+Au 200 GeV collisions at 3 different centralities: $c=25\%, 36\%, 46\%$. Our simulation results (black symbols), by using the PICR hydrodynamic model, shows good agreement with the STAR's experimental data (red and blue symbols). 
}
\label{F1}
\end{figure}

Based on the simulations to the RHIC's Au+Au collisions at the energy $\sqrt s_{NN}= 200$ GeV, we calculate the $\Lambda$ polarization with the widely adopted formula from \cite{BCW2013}.

\bea
\label{Pipv}
 && \vec{\Pi}(p) = \frac{\hbar \varepsilon}{4m}
  \frac{\int \di \Sigma_\lambda p^\lambda \, n_F\ 
  (\nabla\times\vec{\beta})}{\int \di \Sigma_\lambda p^\lambda \,n_F} 
  \nonumber \\
 && + \frac{\hbar {\vec p}}{4m} \times 
  \frac{\int \di \Sigma_\lambda p^\lambda \,
 n_F\ (\partial_t \vec{\beta} + \nabla\beta^0)}
 {\int \di \Sigma_\lambda p^\lambda \,n_F} \ ,
\eea
where $\beta^{\mu}(x)=(\beta^0,\vec{\beta})=[1/T(x)]u^{\mu}(x)$
is the inverse temperature four-vector field, and 
$n_F(x,p)$ is the Fermi-J\"uttner distribution of the $\Lambda$, 
that is $1/(e^{\beta(x)\cdot{p}-\xi(x)}+1)$, being 
$\xi(x)=\mu(x)/T(x)$ with $\mu$ being the
$\Lambda$'s chemical potential and $p$ its four-momentum.  
$\di \Sigma_\lambda$ is the freeze out hyper-surface element, for
$t=$const. freeze-out, 
$\di \Sigma_\lambda p^\lambda\ \to \di V \varepsilon$,
where $\varepsilon = p^0$ being the $\Lambda$'s energy.

The formula indicates that the polarization originates from the relativistic 
thermal vorticity defined as:
\be
\overline{\omega}_{\mu \nu} =
\frac{1}{2}(\partial_\nu\beta_\mu-\partial_\mu\beta_\nu), 
\ee
and is proportional to the mean spin vector:
\be
\vec{S}=\frac{\hbar}{4m}(\varepsilon \vec{\omega}+\vec{p}\times \vec{\omega_0})
\ee
where $\vec{\omega_0}=\overline{\omega}_{0 j}=
\frac12(\partial_t \vec{\beta} + \nabla\beta^0)$ is the temporal component of 
relativistic thermal vorticity, and 
$\vec{\omega}=\overline{\omega}_{ij}=\frac12 (\nabla\times\vec{\beta})$ (i, j =x, y, z) is the spatial component. 
Therefore the polarization vector is just the normalized spin vector $\vec{S}$ by weighting over the $\Lambda$'s number density on the freeze-out surface. E.g. the mean spin vector projected on $y$ direction is,
\be
S_y=\frac{\hbar}{4m}(\varepsilon \omega_{xz}+ (p_x \omega_{0z}+p_z \omega_{0x}))\ .
\ee
Besides, the polarization vector defined in eq. (\ref{Pipv}) is divided into 2 terms:
the {\it first term} arises from the spatial component of relativistic thermal vorticity, 
and the {\it second term} is relativistic modification from the temporal component.

The $\Lambda$ polarization is determined by measuring the
angular distribution of the decay protons in the $\Lambda$'s rest 
frame. In this frame 
the $\Lambda$ polarization is $\vec \Pi_0(\vec p)$, which can be 
obtained by boosting the polarization $\vec \Pi(\vec p)$
from the participant frame to the  $\Lambda$'s rest frame, 
\cite{BCW2013},
\be
\vec{\Pi}_0(\vec{p})=
\vec{\Pi}( p )-\frac
{\vec{p}}
{p^0 (p^0 + m)}
\vec{\Pi}( p ) \cdot \vec{p} \ .
\label{Pi0}
\ee

Finally, since the experimental results for $\Lambda$ polarization are
averaged polarizations over the $\Lambda$ momentum, we evaluated the
average of the $y$ component of the polarization 
$\langle\Pi_{0y}\rangle_{p}$.
We integrated the  $y$ component of the obtained
polarization, $\Pi_{0y}$, over the momentum space as follows:
\ba
\langle\Pi_{0y}\rangle_{p}&=&
\frac{\int dp\, dx \,\Pi_{0y}(p,x) \, n_F(x,p)}{\int dp\, dx\, n_F(x,p)} 
\nonumber \\
&=&\frac{\int \,dp \,\Pi_{0y}(p)\, n_F(p) }{ \int \,dp \,n_F(p)} \ .
\ea

\section{Results and Discussion}

We calculate the $\Lambda$ polarization at 3 impact parameter ratios: $b_0=b/b_{max}= 0.5, 0.6, 0.68$, (where $b$ is the impact parameter and $b_{max}$ is the maximum impact parameter),   corresponding to 3 centrality points: 25\%, 36\%, 46\%. The freeze-out time is 4+4.91 fm/c, except for the 46\% case the FZ time is shorter, i.e. 3.5+4.75fm/c (since smaller system is usually assumed to has shorter evolution time). 

Fig. \ref{F1} shows the centrality dependence of the $y$-directed polarization boosted into $\Lambda$'s rest frame, $\Pi_{0y}$. As expected, the $y$-directed polarization, $\Pi_{0y}$, increases with increasing centrality, since it is already known that the polarization arises from the initial angular momentum, which are related to the impact parameter. This nearly linear dependence on the impact parameter/ centrality were already shown in our previous work
\cite{Xie2017}. 
Fig. (\ref{F1}) is to show that our results (black symbols) of the global polarization for Au+Au 200 GeV, are within the boundary of the STAR data (red and blue symbols), exhibiting a good agreement between them.

\begin{figure}[ht] 
\begin{center}
      \includegraphics[width=7.5cm]{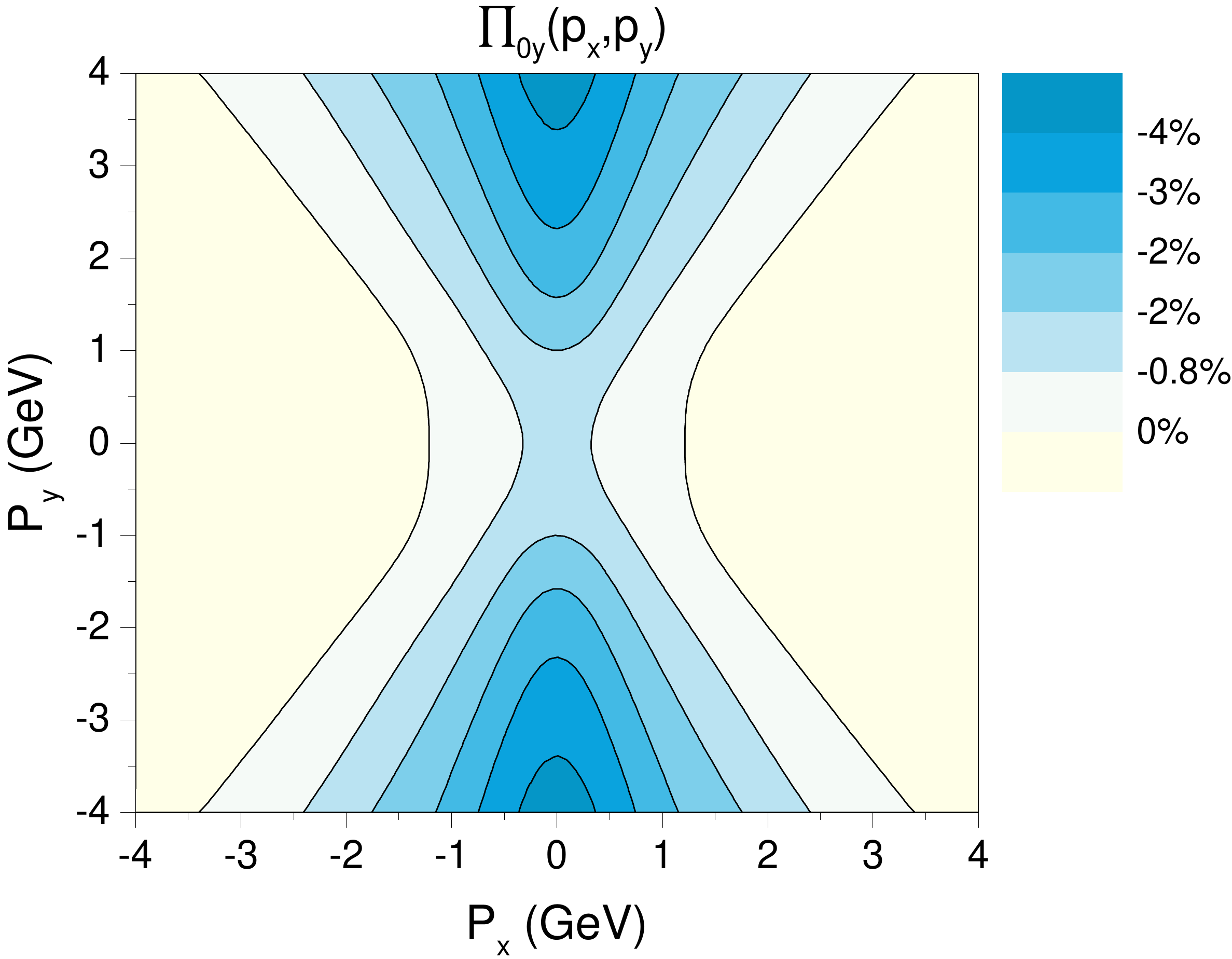}
      \includegraphics[width=7.5cm]{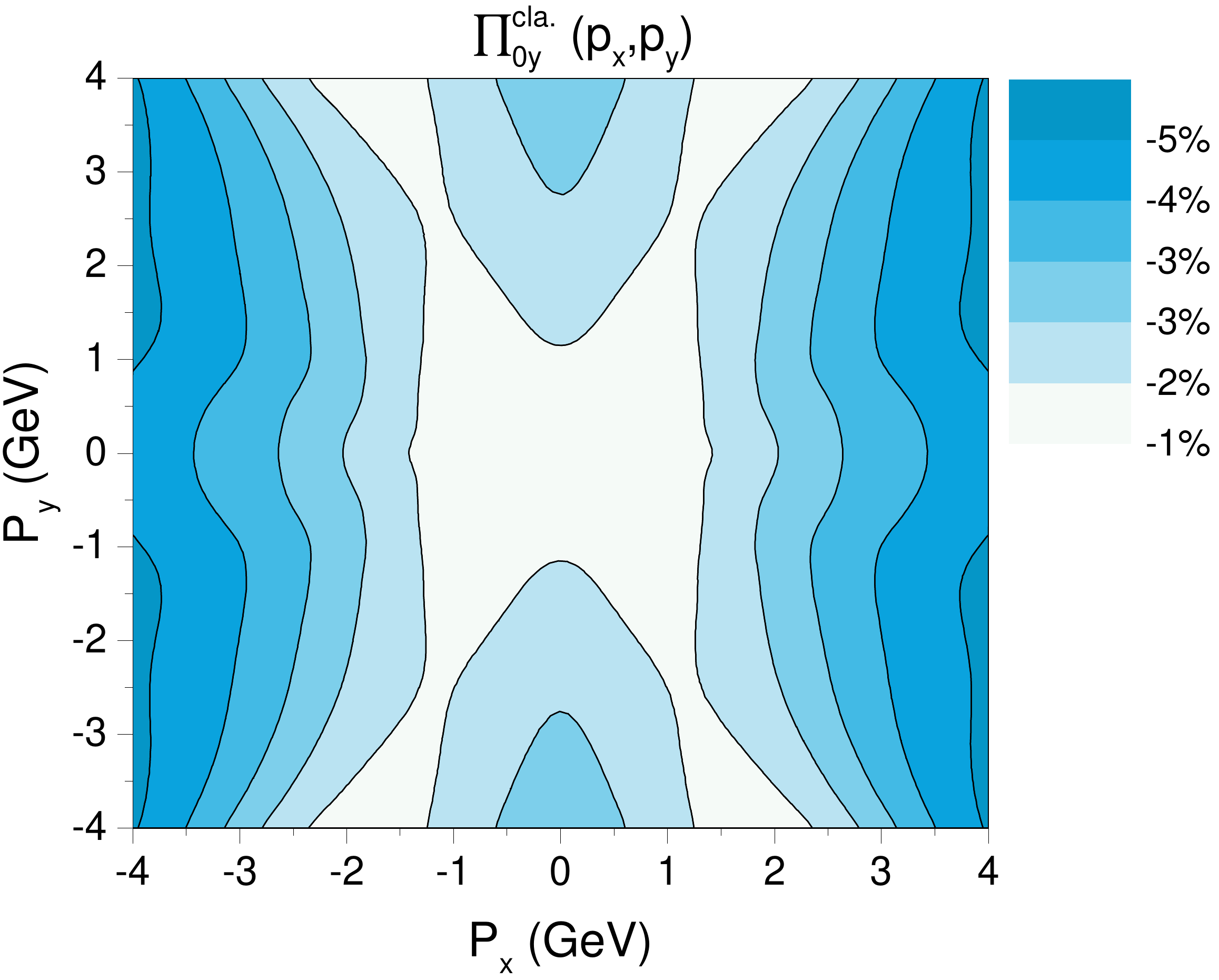}
\end{center}
\caption{
(Color online) The $y$ component of total polarization in the $\Lambda$'s rest frame (upper panel), and of the polarization when only considering the classic vorticity term (lower panel), for Au+Au 200 GeV collisions with impact parameter ratio $b_0 = 0.6$ at the rapidity bin $|y|<1$. 
}
\label{F2}
\end{figure}

Fig. \ref{F2}  (upper panel) shows the $y$ component of polarization in the $\Lambda$'s rest frame, for Au-Au 200 GeV collisions with impact parameter ratio $b_0 = 0.6$ at the rapidity bin $|y|<1$. One can see that the $y$-directed global polarization in $\Lambda$'s rest frame, increases in magnitude from in-plane to out-of-plane, contradicting experimental data. This increasing tendency is the same as our previous results for FAIR's U+U 8GeV collisions
\cite{Xie2016}, 
as well as the other model results
\cite{KB2017,BCW2013}. 
However, as suggested in refs.
\cite{ITS2019,FKR2019}, 
we also calculate the global polarization by only considering the first term of the polarization vector in eq. (\ref{Pipv}) and boosting it into $\Lambda$'s rest frame, as shown in down panel of Fig. \ref{F2}. One can see that the polarization decreases from in-plane to out-of-plane, agreeing with the experimental data, whose magnitude, however, is about ten times smaller. Actually this is nothing new, but had been shown in our previous work 
\cite{BCW2013}.

\begin{figure}[ht] 
\begin{center}
      \includegraphics[width=8.5cm]{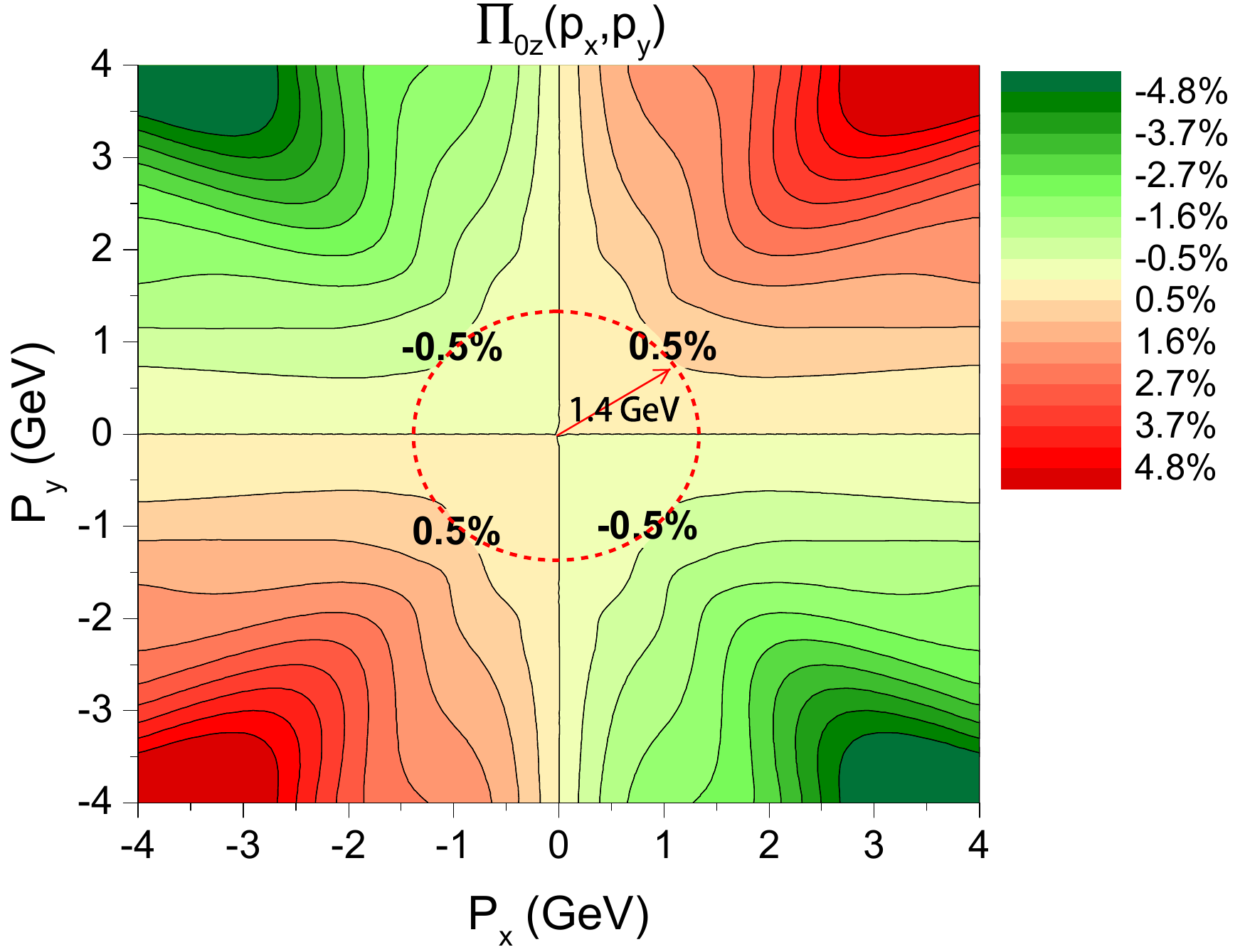}
\end{center}
\caption{
(Color online) The transverse momentum distribution of longitudinal polarization, $\Pi_{0z}$, for Au-Au 200 GeV collisions with impact parameter ratio $b_0=0.68$ at rapidity bin $| y | < 1$. 
}
\label{F3}
\end{figure}

Fig. \ref{F3} shows the transverse momentum distribution of longitudinal polarization, $\Pi_{0z}$,   with impact parameter ratio $b_0=0.68$ for the 200GeV Au+Au collisions. It has the correct sign distribution compared to the experimental data, which is (+, - , +, -) counting from the first coordinate quadrant to fourth quadrant. The peak value at $p_T = 1.4$  GeV is about 0.5\%, the same as the global polarization at $b=0.5$ fm/c. This is in agreement with the experimental data, i.e. the peak value of longitudinal polarization at $p_T = 1.4 $ GeV has similar magnitude with the global polarization.

\begin{figure}[ht] 
\begin{center}
      \includegraphics[width=8.5cm]{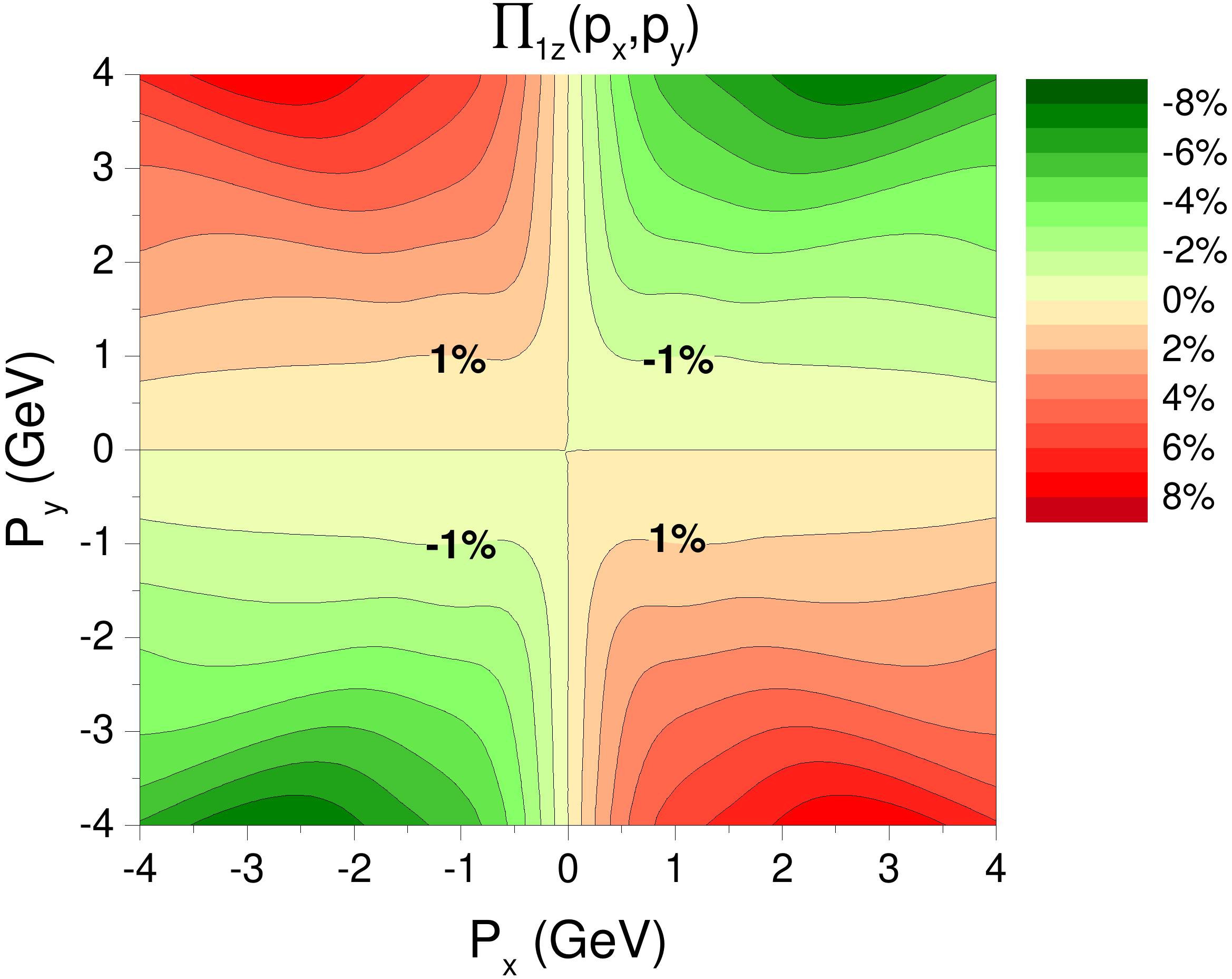}
      \includegraphics[width=8.5cm]{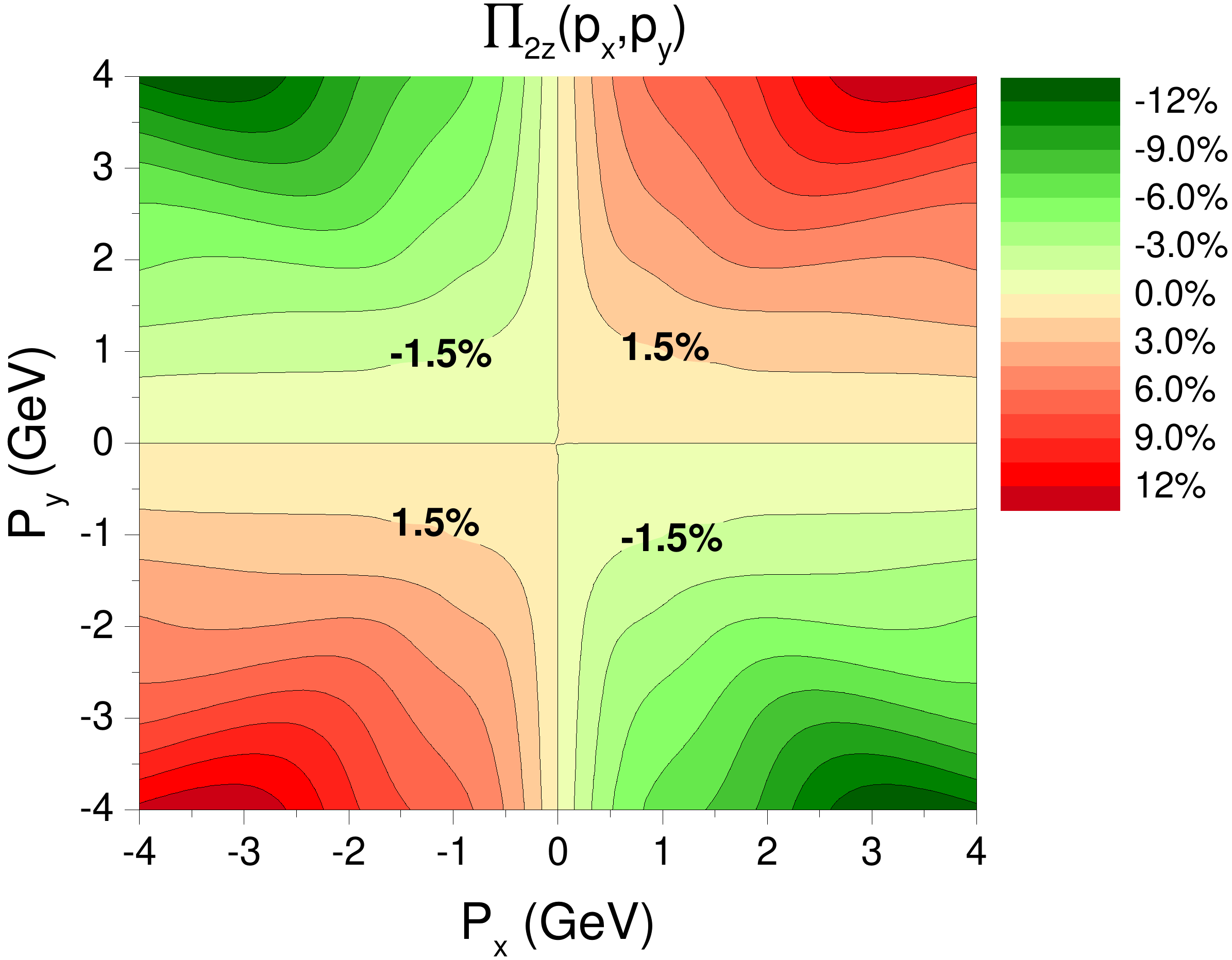}
\end{center}
\caption{
(Color online) The first term of longitudinal polarization, $\Pi_{1z}$,  and sencond term of longitudinal polarization, $\Pi_{2z}$, distributed on transverse momentum plane, for Au+Au 200GeV collisions with impact parameter ratio $b_0=0.68$ at rapidity bin $| y | < 1$. 
}
\label{F4}
\end{figure}

Furthermore, as shown in Fig. \ref{F4} we find that the first term has a sign distribution (-, + , -, +),  but the second term has the opposite signature and a larger magnitude, resulting in a (+, - , +, -)  sign distribution that agrees with experimental data. Comparing to our previous results for FAIR's U+U 8GeV collisions 
\cite{Xie2016}, 
the first term keeps the same sign distribution, i.e. (-, + , -, +), but with magnitude growing from about 2\% to 8\% at large transverse momentum. Meanwhile, the second term flips it sign distribution, from (-,+ ,-,+) to (+, - , +, -), and grows faster to a magnitude of 12\%, which is larger than the first term. Two points are worthy to be noticed here:

(1) The magnitude, of either the first/second term or the total of longitudinal polarization, increases from low energy (8 GeV) to high energy (200 GeV). This seems contradicts with a previous work 
\cite{BK2018}, 
where the second harmonic coefficient  of the longitudinal polarization decreases with energy increasing from 7.7 GeV to 2.76 TeV;
 
(2) The second term, in our model, plays crucial role to obtain the experimentally observed sign structure and magnitude of the longitudinal polarization: it has a sign structure of (+, - , +, -), and a larger magnitude, covering the first term's opposite signature and amending the polarization value into a smaller but correct magnitude. This is similar to ref. 
\cite{FKR2019}, 
where the total longitudinal polarization flips its sign distribution with respect to that of the first term, although the signatures therein are just opposite to our results.

\begin{figure}[ht] 
\begin{center}
      \includegraphics[width=8.5cm]{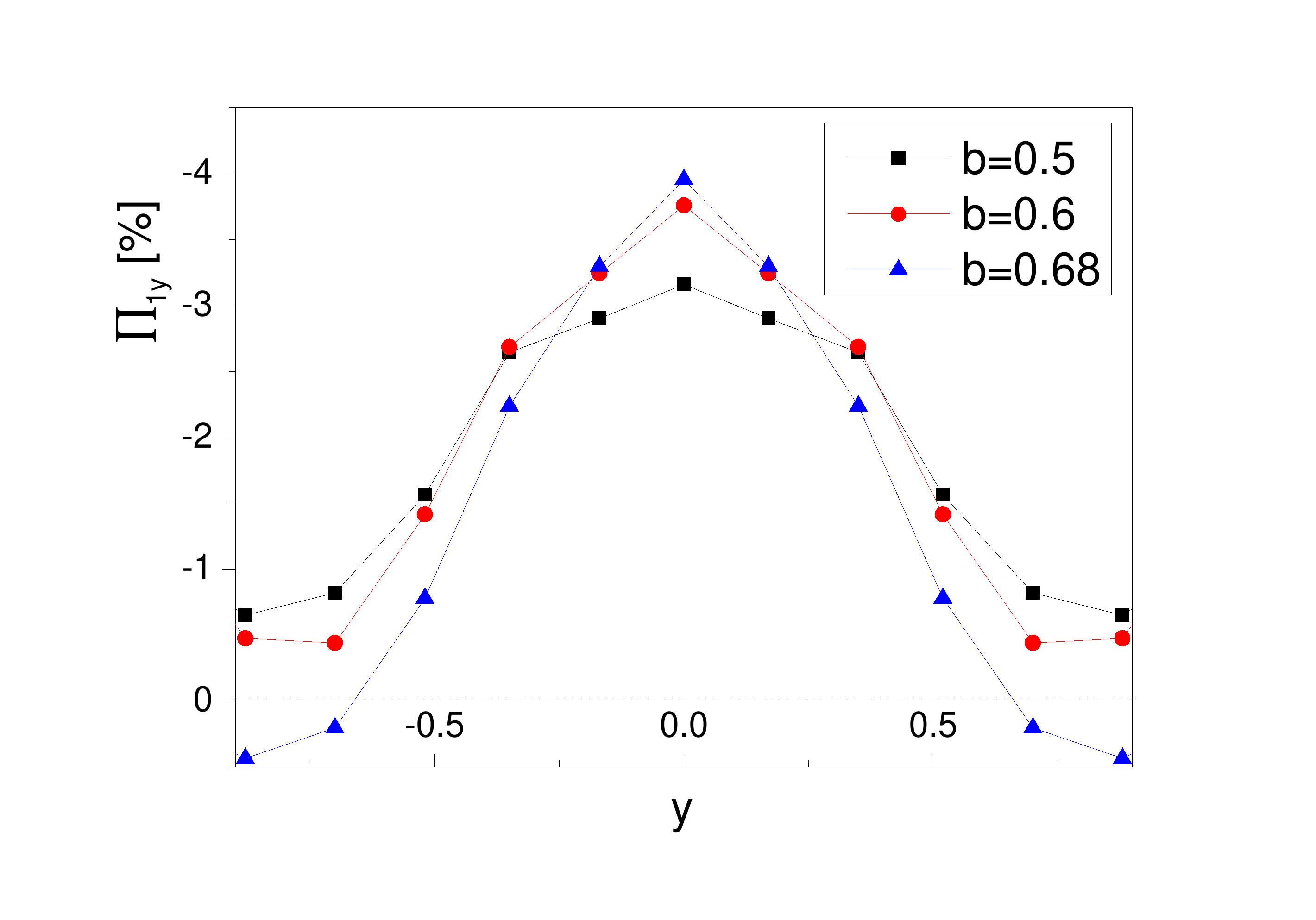}
      \includegraphics[width=8.5cm]{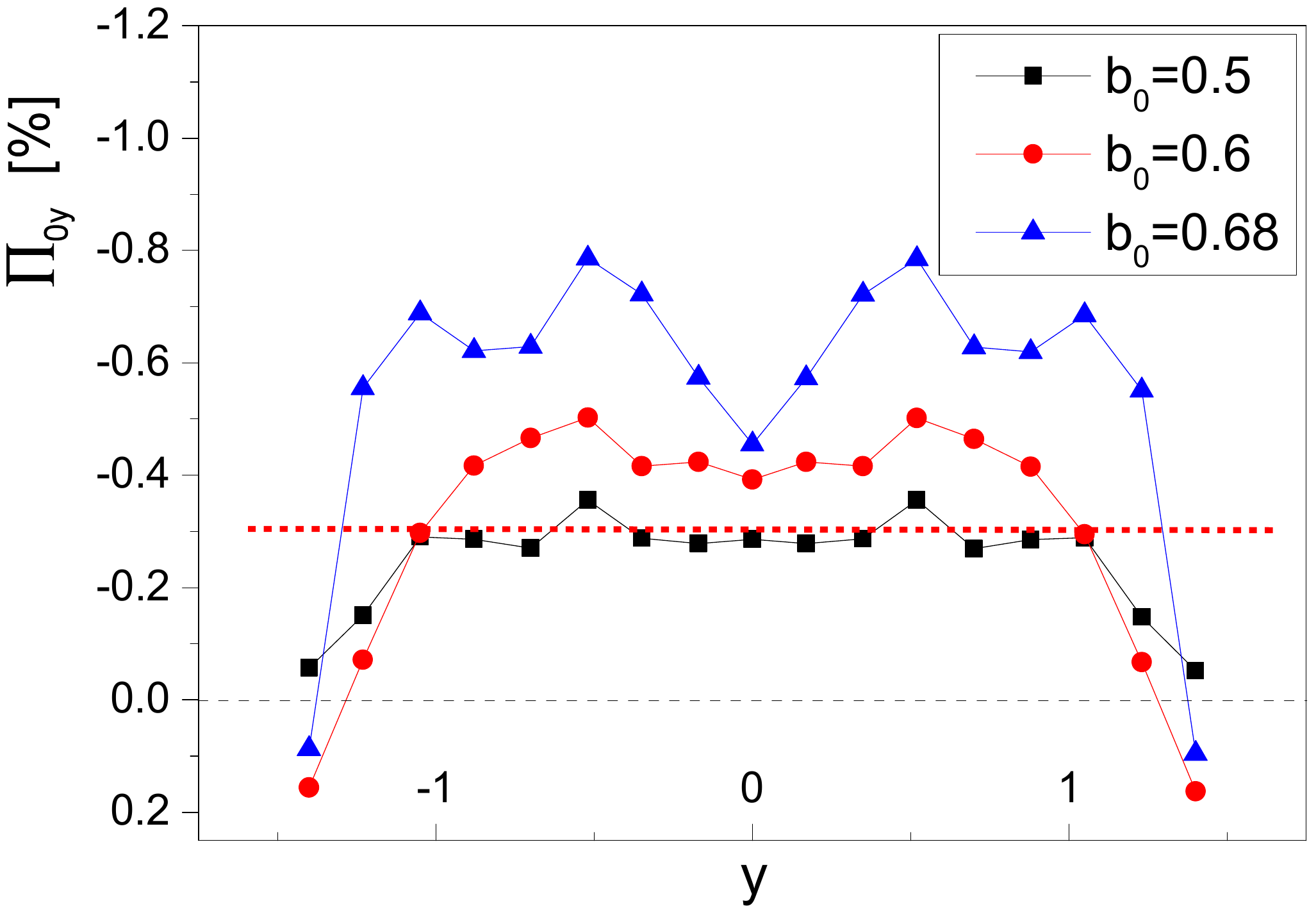}
\end{center}
\caption{
(Color online)
The first term of the polarization (upper panel), and the global polarization (lower panel) as a function of rapidity at different impact parameter ratios. The red dashed line in the lower panel figure is a rough approximation of the experimental data. 
}
\label{F5}
\end{figure}

Then we explore also the global polarization as a function of rapidity, as shown in the lower panel of Fig. \ref{F5}.
The red dashed line in the lower panel figure is a rough approximation of the experimental data, which shows no significant dependence on the rapidity and fluctuates around the averaged value 3\%. One can see that the global polarization from our model also shows no significant dependence on the rapidity. The global polarization, $\Pi_{0y}$, for $b_0 = 0.5, 0.6 ,0.68$, fluctuates around the average value of 2.8\%, 3.8\% and 6\% respectively, which are magnitudes similar to the global polarization. 
For more peripheral collisions, the fluctuations are relatively larger, e.g. at the case of $b_0 = 0.68$, there exists a dip in rapidity bin $|y| < 0.4$. Beyond the rapidity range $|y| > 1$ the global polarization goes down rapidly to zero.

The first term of the $y$-directed polarization, as shown in the upper panel of Fig. \ref{F5}, exhibits a normal distribution with respect to the rapidity, with peak value at center rapidity $y = 0$, which is similar to the vorticity distribution on pseudo-rapidity from AMPT model\cite{DH2016}. This similarity of structure simply demonstrates the definition of polarization vector's first term, i.e. $\Pi_{1y}$ arises purely from the spatial component of relativistic vorticity,  $\vec{\omega} =\frac12 \nabla \times \vec{\beta}$. For more peripheral collisions with larger impact parameter, the global polarization distribution peaks higher at center rapidity $y = 0$ and  goes down faster to zero with a narrower width. Finally, the two figures together indicate that the second term related to the system expansion, flatten the peak of the first term induced by classical vorticity, resulting in an even distribution of global polarization on the rapidity.

\section{Summary and Conclusions}

With a Yang-Mills field stratified shear flow initial state and a high resolution (3+1)D Particle-in-Cell Relativistic (PICR) hydrodynamic model, we calculate the $\Lambda$ polarization for Au-Au collisions at RHIC energy of $\sqrt{S_{NN}}=200$ GeV. The transverse momentum distribution of global polarization shows a magnitude increasing from in-plane to out-of-plane, contradicting to the experimental data. However, the longitudinal polarization in our model shows correct signature of the experimentally observed quadrupole structure on transverse momentum plane. Besides, the peak value of the longitudinal polarization at $p_t =1.4$ GeV is similar to the global polarization, which is in good agreement with the experimental data. When delving into the two terms of the polarization vector, it is found that the second term arising from system expansion or the temporal component of the relativistic thermal vorticity, plays a crucial role to obtain our results, by changing the signature and magnitude of the first term which is induced by classic vorticity. Furthermore, we also plot the global polarization as a function of rapidity. Our results show no significant dependence on rapidity, which again conform with the experimental data, and again it is found that it is the second term that flattens the sharp peak of the first term induced by classical vorticity.

\section*{Acknowledgments}

We thank  Che Ming Ko and Gang Chen for enlightened discussions. The work of L. Cs. is supported by the Research Council of Norway Grant No. 255253/F50, and of Y. L. Xie is supported by the Fundamental Research Funds for the Central Universities (Grant No. G1323519234).




%
%

\end{document}